\documentclass[aps,prl,twocolumn,showpacs,floatfix,superscriptaddress,nofootinbib]{revtex4}
\usepackage{graphicx}
\usepackage{hyperref}
\usepackage{amsmath}
\usepackage{times}
\newcommand{\ket}[1]{|#1\rangle}

\def\>{\rangle}
\def\<{\langle}

\begin{document}

\title{Loss tolerance in one-way quantum computation via counterfactual error correction}

\author{Michael Varnava}
\address{QOLS, Blackett Laboratory, Imperial College London, Prince Consort Road, London SW7 2BW,
UK}

\author{Daniel E. Browne}
\address{Department of Materials and Department of Physics, University of Oxford, Parks Road, Oxford, OX1 3PU, UK}

\author{Terry Rudolph}
\address{QOLS, Blackett  Laboratory, Imperial College London, Prince Consort Road, London SW7 2BW,
UK}
\address{Institute for Mathematical Sciences, Imperial College London, 53 Exhibition Road,
London SW7 2BW, UK}

\begin{abstract}
We introduce a scheme for fault tolerantly dealing with losses (or other ``leakage'' errors) in cluster state computation that can tolerate up to 50\% qubit loss.
This is achieved passively using an adaptive strategy of measurement - no coherent measurements or coherent correction is required. Since the scheme relies on inferring information about what would have been the outcome of a measurement had one been able to carry it out, we call this  \emph{counterfactual} error correction.

\end{abstract}


\maketitle

Quantum computation architectures can only be considered viable if
they are demonstrably fault tolerant. Thresholds on fault tolerance
are typically quoted as around $0.01\%$ - that is, the error on a
generic gate operation or state preparation should be less than this
amount. However, one hopes that for specific error models well
chosen strategies might significantly relax this rather stringent
requirement \cite{knill05}.  

A new architecture for quantum computation that is generating much interest is the \emph{cluster state} model \cite{clusterqc1,clusterqc2,danreview}, otherwise known as ``one-way quantum computation''. In this approach, the quantum computation proceeds in two stages. First, a special kind of entangled multi-party state, called a cluster state is generated. The quantum computation is then implemented by single qubit measurements; the specific algorithm computed is a function of the choice of measurement bases, and the order in which they are made. There are two main approaches to tackling fault tolerance in a new architecture. One is to translate standard fault tolerance techniques to the new model, an approach which has been successful for the cluster state quantum computation \cite{raussenthesis,dawson,aliferis}. A different approach is to develop novel protocols which exploit the features of the architecture, in this case the entanglement of the cluster states themselves. This is the approach we have adopted here. A further recent development in this vein is a fault-tolerant model which exploits the fact that certain cluster states are topological error correcting codes \cite{raussen}.

A main result of this paper will pertain to architectures which can
produce and utilize cluster states in such a way that a significant
error mechanism involves each qubit in the cluster being lost (or
undergoing some other sort of detectable failure) with fixed and
independent probability $\varepsilon_0$. What we will show is
\begin{quote}
If $\varepsilon_0<50\%$ then computation fault tolerant to loss errors is possible.
\end{quote}
Remarkably this can be achieved with only destructive measurements,
and without coherent correction mechanisms needing to be applied to
the qubits.

In a subsequent paper \cite{mikedantezloqcloss}, we will show how this result can be exploited to allow efficient linear optical quantum computation whenever the product of source and detector efficiency is greater than 2/3, a dramatic improvement on previous thresholds (preliminary results can be found in \cite{preprint}).

\begin{figure}
\begin{center}\includegraphics[width=6cm]{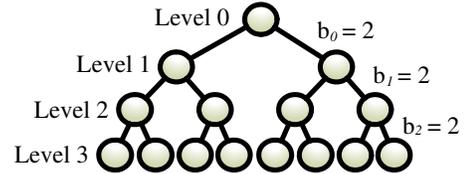}\end{center}
\caption{ \label{tree} A tree cluster with branching
parameters $b_{{0}}=b_{{1}}=b_{{2}}=2$.}
\end{figure}

\begin{figure}
\begin{center}\includegraphics[width=6.3cm]{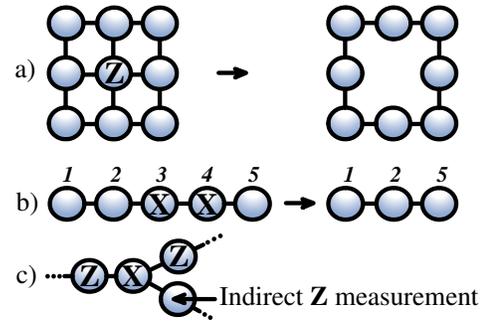}\end{center}
\caption{\label{cproperty}Certain measurements on a graph qubit
 leave the remaining qubits in a new cluster state with a
different layout: a) A $Z$ eigenbasis measurement removes the
qubit from the cluster and breaks all bonds between that qubit and
the rest. b) Two adjacent $X$ measurements on
a linear cluster remove the qubits and form
direct bonds between their neighbors. c) Measuring a qubit
in the $X$ basis and performing $Z$ measurements on
all but one of the remaining directly bonded qubits
deterministically reveals what the $Z$ measurement outcome
would have been on the unmeasured qubit.}
\end{figure}


We begin by introducing cluster state structures for
quantum computation which are naturally resilient to qubit loss. 
To simplify the first part of our discussion, we shall initially assume that any loss errors occur after the entangled state has been generated. Later on, we shall relax this requirement.

We
use  $X_i$,   $Y_i$, $Z_i$ to denote the Pauli operators
on qubit $i$, and define cluster states in terms of
their stabilizers \cite{clusterqc2} as follows: A cluster state
is represented by a graph, where the $n$ vertices of
the graph denote qubits, while the bonds denote a certain entangling
operation between them. 
Denoting by $E(i)$ the set of edges on this
underlying graph which are connected to vertex $i$, the cluster state is that state which is invariant under the action of
the $n$ stabilizer operators:
\[
X_i \prod_{j\in E(i)} Z_j.
\]

Equivalently the cluster state is the state obtained by preparing each qubit in the state $\ket{0}+\ket{1}$ and then applying a CPHASE (controlled phase-flip) gate to every pair of qubits connected by an edge on the graph.

Since the cluster states are eigenstates of the stabilizers, they predict with certainty correlations in the measurement
outcomes of certain sets of measurements. To illustrate this,
consider a two-qubit state stabilized by the operator $X_1 Z_2$. If
observable $X_1$ is then measured the outcome of $Z_2$ is now known
with certainty. We say that the measurement $X_1$ is an
\emph{indirect measurement} of the observable $Z_2$. Importantly, if
such a state is prepared, and qubit 2 undergoes a loss error, $Z_2$ can still
be measured indirectly, even though qubit 2 is no longer available.
The principle of applying counterfactual reasoning  to the correlations in cluster states to effect indirect measurements on lost qubits, lies at the heart of our
 scheme.

\begin{figure}
\begin{center}\includegraphics[width=5cm]{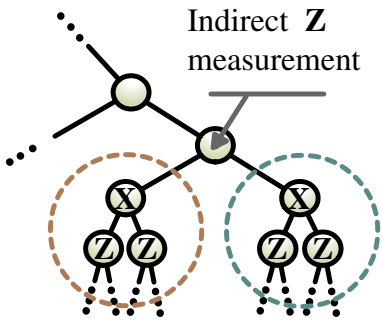}\end{center}
\caption{ \label{indirect} An indirect $Z$ measurement will be
performed on the indicated qubit if either the three measurements
circled in red are successful, or if those circled in green are
successful. }
\end{figure}

The cluster state structure we utilize is that
of a tree (see Fig.~\ref{tree}), which can be specified
by its \emph{branching parameters}, $b_{\emph{i}}$ as indicated.
The tree graphs are designed in this way to take full advantage of
the three general cluster state properties summarized in
Fig.~\ref{cproperty}. In particular, a crucial feature of these tree clusters is that
any given qubit within the cluster can be removed indirectly
(utilizing an effective $Z$ measurement) by performing measurements
on a subset of the qubits below it in the cluster. This is depicted in
Fig.~\ref{indirect}.

\begin{figure}
\begin{center}\includegraphics[width=7cm]{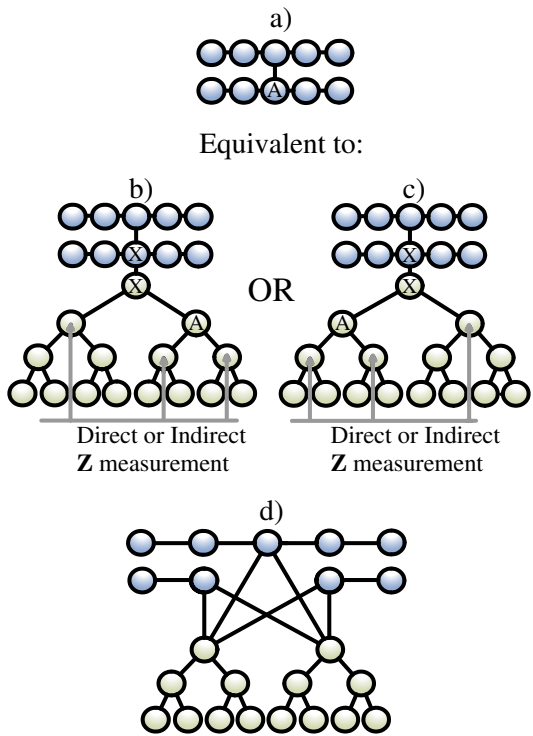}\end{center}
\caption{ \label{treeappl} A standard cluster computation requires measurement of observable $A$ on the indicated qubit, as indicated in (a). Because of the properties depicted in Fig.~\ref{cproperty}, either of the two measurement patterns depicted in (b) and (c) will implement the desired measurement of $A$. In (d) we see the cluster state that is generated on the remaining qubits when the two $X$ measurements marked in (b) and (c) are made. This illustrates that these two qubits do not  exist in the final loss tolerant graph state. They are, however, a useful didactic tool to understand the scheme, and they may be used in the probabilistic generation of the final state.
}
\end{figure}

Suppose we desire to perform a quantum computation using the cluster
shown in Fig.~\ref{treeappl}(a). In general, the computation
requires the measurement of some observable, $A(\alpha)\equiv\cos\alpha
X+\sin\alpha Y$, directly on a given qubit as indicated. However,
due to the properties (b) and (c) of Fig.~\ref{cproperty}, it is
clear that an equivalent measurement procedure is given by
Fig.~\ref{treeappl}(b). The important point to note, is that there
is an alternative measurement pattern - that of
Fig.~\ref{treeappl}(c) - which also achieves exactly the same
effect. Now, since they are in the Pauli group, the pair of $X$
measurements can be performed
`offline', i.e. as part of preparing the cluster state for the
computation and prior to the computation having reached this
position. (Alternatively the cluster state in Fig.~\ref{treeappl}(d) which results after these $X$-measurements could be grown directly).

The core of our loss tolerance results from the fact that
first the $A$ measurement can be attempted as in Fig.~\ref{treeappl}(b).
If our measurement device fails to register an outcome, however, due, for example, to a loss error,  then we try and remove this
qubit of the cluster by an indirect $Z$ measurement, and proceed to make the $A$
measurement as in Fig.~\ref{treeappl}(c). If this measurement also
fails (or if any of the necessary $Z$ measurements fail) then the
computation has failed. As we now show, the generalization of this
simple procedure to a tree with larger branching parameters can greatly reduce the effective qubit loss rate and lead to a successful implementation of the $A$ measurement. 

Note that the order of the measurements demanded by the our loss-tolerance protocol (as described in the previous paragraph) is not the one that would allow the implementation of a deterministic rotation gate. In fact the sign of this rotation angle depends on the outcome of
measurements which must (for loss tolerance) be made after $A$ itself is measured.
Thus with equal probability a rotation corresponding to either $A(\alpha)$  or $A(-\alpha)$ occurs.
Fortunately, by adopting strategies described in
\cite{nielsenmeas} this effect can be overcome at the cost of a small
overhead in the size of the computation. Note that the measurement order imposed in the equivalent  one-way quantum computation without loss tolerant encoding (as described in the introductory paragraphs) must still be respected, as otherwise one could not determine whether  $A(\alpha)$  or $A(-\alpha)$ had been implemented and thus which corresponding correction strategy one should employ.

Consider a tree cluster with branching parameters
$b_{{0}},b_{{1}},b_{{2}},.....b_{\emph{m}}$. The probability of successfully performing the necessary
measurement pattern to implement the $A$ measurement on such a
tree is given by
\begin{equation}\label{Psucc}
P=[(1-\varepsilon_0+\varepsilon_0
R_{{1}})^{b_{{0}}}-(\varepsilon_0
R_{{1}}){}^{b_{{0}}}](1-\varepsilon_0+\varepsilon_0
R_{{2}})^{b_{{1}}},
\end{equation}
where for $k\le m$
\begin{equation}\label{Reqs}
R_{\emph{k}} =1-[1-(1-\varepsilon_0)(1-\varepsilon_0+\varepsilon_0 R_{{k+2}})^{b_{{k+1}}}]^{b_{\emph{k}}}
\end{equation}
and $R_{{m+1}}\equiv 0,b_{m+1} \equiv 0.$\\


The $R_\emph{i}$ are defined as the probabilities of successfully
implementing an indirect $Z$ measurement on any given qubit found in
the $i^{th}$ level. Thus $(1-\varepsilon_0)+\varepsilon_0 R_\emph{i}$ is
the probability of a successful $Z$ measurement on a qubit at level
$i$. The equations (\ref{Reqs}) can be derived recursively, starting from the qubits at the  bottom of the tree.

The derivation of $P$ can be understood as follows. Imagine one
proceeds along the $b_{0}$ qubits until the $A$ measurement
succeeds on qubit number $k$. Overall success, given that the $A$
measurement has succeeded on qubit $k$, requires: (i) Successful
(direct or indirect) $Z$ measurements on the $b_{1}$ qubits
below qubit $k$, (ii) indirect Z measurements to succeed on the
$k-1$ qubits that the $A$ measurement failed on and (iii) direct or
indirect $Z$ measurements to succeed on the remaining $b_{0}-k$
qubits. This leads to an overall success probability $P$. 

\begin{figure}
\begin{center}\includegraphics[width=9cm]{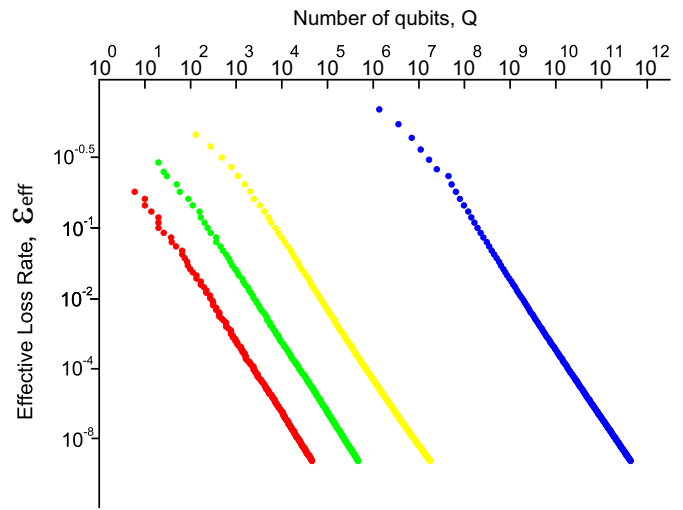}\end{center}
\caption{ \label{numerfig} Numerical results showing the the number of qubits required versus the desired effective loss rate.  The different curves, from left to right, correspond to $\varepsilon_0=0.2,0.3,0.4,0.49$ respectively.}
\end{figure}


Let us introduce the parameter $\varepsilon_\textrm{eff}\equiv1-P$ to represent the overall effective loss rate for the tree-encoded logical cluster qubit.
 For a range
of fixed $\varepsilon_0$ and $\varepsilon_\textrm{eff}$ we have performed a numerical  search over
integer values for the branching parameters $b_{0},b_{1},\ldots
b_\emph{m}$ which minimize the number of qubits $Q$ required in the
tree. These results are depicted in Fig.~\ref{numerfig}. (Note the
plot is log versus loglog.) We have shown the abscissa to somewhat
extreme values only to indicate the important scaling. The linear nature of the plot implies that
$Q=\textrm{polylog}\left(1/{\varepsilon_\textrm{eff}}\right)$, which demonstrates that this
procedure has the scaling properties required for proper fault
tolerance.

When $\varepsilon_0\ge0.5$, numerical simulations show that
$\varepsilon_\textrm{eff}$ cannot be lowered from $\epsilon_0$. This indicates a loss tolerance threshold of 50\%. This can be
understood as follows. If the above-described scheme could correct
loss errors with an error rate at 50\% or higher, this would violate
the no-cloning theorem \cite{noclone}. To see this, consider a  4-qubit linear cluster state. Alice is given the first three qubits, and Bob is given the last, which Bob then immediately encodes  in a tree. Alice makes measurements with bases unknown to Bob.
 Now, imagine a third party, Charlie, steals 50\% of Bob's qubits, chosen at random. This would be  indistinguishable (to Bob) from the qubits having a 50\% loss rate. If our protocol (with a non-demolition measurement replacing the measurement in basis ``A'') allowed this level of loss to be tolerated, both Bob and Charlie could produce ``clones'' of the state prepared by Alice's measurements - violating the no-cloning theorem.

Interestingly, by allowing an adaptive construction of the tree based on which particular
qubits in the first row are lost, the no cloning theorem is no
longer an obstacle and a strategy can be found for
 arbitrarily high loss rates \cite{tomprivcom}.
This strategy needs deterministic gate operations
to adaptively build up the tree, which makes it incompatible with
non-deterministic cluster state generation schemes \cite{nielsencluster,efficientTezDan,BK,Kieling}.

So far, our error model has only explicitly considered loss errors occurring after the cluster state has been generated. Fortunately, we can relax that requirement in a number of practically relevant scenarios. First let us consider cluster state generation via deterministic CPHASE gates.
Loss errors can occur \emph{before}  the CPHASE without disturbing the efficacy of the loss tolerance protocol if the attempted physical entangling operation acts (when only one physical qubit carrier is present)  as the identity (up to a global phase).  The loss error takes the state outside of the computational basis and so the action of an abstract CPHASE gate is not defined. However, for any physical implementation of such an operation, this can be assessed and, indeed,  a number of proposed quantum gates,  e.g. \cite{quantumgates} satisfy this criterion.

To confirm the loss correction protocol also succeeds in this case,  recall that obtaining the +1 outcome of a  $Z$ measurement on a given qubit in the cluster leaves the remaining qubits in  a  cluster state with the same graph, but  with all edges attached to the measured qubit removed (as illustrated in Fig.~\ref{cproperty}~(a). This, however, is the same state which would have been obtained if no entangling operations on that qubit had  been applied or, equivalently, if that qubit had suffered a loss error before the operations.  Thus in this case  loss events before the entangling operation are equivalent to loss events afterward, and will be corrected in the same way.

In linear optical quantum computation, deterministic CPHASE gates are not available. It is possible to combine  non-deterministic gates with a near-deterministic gate teleportation approach \cite{klm}, or, more efficiently, generate the cluster state via non-deterministic gates and an appropriate strategy \cite{nielsencluster,efficientTezDan,BK}. However, the most resource efficient approach is to employ so-called non-deterministic ``fusion operations'' \cite{efficientTezDan} via polarizing beams splitter networks. This setup requires a more complicated analysis, but in this case also, loss errors occurring before and during the beam splitter network can still be tolerated. For a full description and proof see \cite{mikedantezloqcloss}.

In practice, of course, a quantum computation must be tolerant to more
than loss errors. In this regard there are several important things to
note about our scheme. Firstly, because of the large overhead in numbers
of qubits in the tree, one might naively expect that the loss tolerance
comes at the expense of creating large sensitivity to depolarizing
noise. In fact this is not the case. For a start, when increasing  the size of the tree, the vast majority of the extra qubits will only ever be measured in a Pauli basis. For example, for an initial loss rate $\varepsilon_0=0.2$, to get
$\varepsilon_\textrm{eff}=10^{-10}$ then non-Pauli measurements will only be attempted on at most  15 of the qubits forming each tree. Note also that the vast majority of operations here  are in the  Clifford group. Robust  fault-tolerant methods of implementing Clifford operations are well-known (see e.g. \cite{nielsenchuang}).

However additional encoding of these operations is not even necessary. There is in fact, a natural robustness against general errors in our loss-tolerance protocol. This is because the tree structure leads to a great deal of redundancy. For every qubit, there are a number of different ways an indirect measurement can be made, e.g. measurements on different subsets of qubits below it on the tree. These different sets of measurements  would, in the ideal case, all lead to the same result, but the presence of additional noise will lead to errors. Erroneous indirect measurements will lead to additional logical errors (both local depolarization and non-Markovian errors \cite{dawson,aliferis}) in the computation. Adopting a simple majority voting strategy, however, significantly decreases the overall error rate. In fact, the greater the branching ratios  in the trees, the lower the probability of an incorrect majority vote.

Qubits and measurements in the first row of the tree (the qubits which may need to be measured in a non-Pauli basis) are not protected from errors (and the negligible residual error caused by incorrect indirect measurements) by such a strategy. However, by encoding these qubits further, using an error-detecting code, more general errors could be detected and located. The protocol can tolerate located errors in the same way as losses. The 15-qubit quantum Reed-Muller code \cite{bravyi} is most suitable for this task \cite{raussen}  as it is the only code known so far,  where the non-Pauli measurements on the encoded qubit can be achieved with single qubit measurements.

Our primary purpose in this paper has been to show that an extremely
high error threshold exists for one of the primary error mechanisms
affecting several quantum computational architectures based on the
one-way model. This is achieved using tree-like cluster state structures to implement counterfactual measurements on qubits which have undergone a loss error. It seems likely to us that practical resource
reductions may be achieved by utilizing different graph states than
the specific tree clusters we have considered, or by using more
dynamical approaches wherein the cluster trees are built as needed
(rather than all a priori as we have considered here). Our scheme will be relevant to all realizations of cluster state quantum computation where qubit
loss is an important source of error,  especially optical realizations.


We acknowledge useful conversations with C.M. Dawson, C. Moura Alves, J. Franson,
T. Jennewein, P. Kok, P. Kwiat, M.A. Nielsen, H. Ollivier, J.-W. Pan, T. Stace, I.
Walmsley, P. Walther, and G. Weihs. This research was supported by
The British Council, Merton College, Oxford, the QIPIRC and the
Engineering and Physical Sciences Research Council.


\begin{thebibliography}{99}
\bibitem{knill05} E. Knill, Nature (London) \textbf{434} 39 (2005); M. Silva, M. Roetteler, C. Zalka,   quant-ph/0502101.

\bibitem{clusterqc1}  R. Raussendorf and H. J. Briegel, Phys. Rev. Lett. {\bf 86}, 5188 (2001).

\bibitem{clusterqc2} R. Raussendorf, D.E.~Browne and H.J.~Briegel, Phys. Rev. A {\bf 68}, 022312 (2003).
\bibitem{danreview} D.E. Browne and H.J. Briegel, quant-ph/0603226.

\bibitem{dawson}  M.A. Nielsen and C.M. Dawson, Phys. Rev. A \textbf{71}, 042323 (2005);  C.M. Dawson, H.L. Haselgrove, M.A. Nielsen, Phys. Rev. Lett. \textbf{96}, 020501 (2006).

\bibitem{raussenthesis}  R. Raussendorf, PhD Thesis, Ludwig-Maximilians University, Munich (2003), online at http://edoc.ub.unimuenchen.
de/archive/00001367/.

\bibitem{aliferis}  P. Aliferis and D.W. Leung, Phys. Rev. A \textbf{73}, 032308 (2006).
\bibitem{raussen} R. Raussendorf, J. Harrington and K. Goyal,  quant-ph/0510135. 


\bibitem{mikedantezloqcloss} M. Varnava, D.E. Browne and T. Rudolph, in prep. (2006). 
\bibitem{preprint} M. Varnava, D.E. Browne and T. Rudolph,   quant-ph/0507036v2. 
\bibitem{nielsenmeas} M.A.~Nielsen, Phys. Lett. A. \textbf{308}, 96 (2003).

\bibitem{noclone} W. K. Wooters and W. H. Zurek, Nature \textbf{299} 802 (1982); D. Diekes, Phys. Lett. A \textbf{76} 271 (1982).
\bibitem{tomprivcom} T. Stace, Private communication (2005).

\bibitem{nielsencluster} M.A. Nielsen,  Phys. Rev. Lett. {\bf 93} 040503 (2004).
\bibitem{efficientTezDan} D.E. Browne and T. Rudolph, Phys. Rev. Lett. \textbf{95} 10501 (2005).
\bibitem{BK} S.D. Barrett and P. Kok,  Phys. Rev. A \textbf{71}, 060310(R) (2005).
\bibitem{Kieling} K. Kieling, D. Gross and J. Eisert,   quant-ph/0601190, D. Gross, K. Kieling and J. Eisert, quant-ph/0605014.


\bibitem{quantumgates}  D. Jaksch, H.J. Briegel, J.I. Cirac, C.W. Gardiner and P. Zoller, Phys. Rev. Lett. 82, 1975 (1999).
\bibitem{klm} E. Knill, R. Laflamme and G. Milburn, Nature (London) \textbf{409}, 46 (2001).





\bibitem{nielsenchuang}  M.A. Nielsen and I. Chuang,
\newblock {\em Quantum Computation and Quantum Information},
 (Cambridge University Press, Cambridge, 2000). 

\bibitem{bravyi} S. Bravyi and A. Kitaev, Phys. Rev. A 71, 022316 (2005).


\end{thebibliography}
\end{document}